\documentclass[aps,prd,preprint,superscriptaddress,tightenlines,nofootinbib]{revtex4}
\usepackage{dcolumn}% Align table columns on decimal point
\usepackage{bm}% bold math
\usepackage{epsfig}

\newcommand{\etal}{{\it et al.}}

\begin{document}

\preprint{CLEO CONF 04-13}   % For conference papers
\preprint{ICHEP04 ABS11-0778}      % For conference papers
%title
\title{\LARGE Evidence for $B_s^{(*)}\overline{B_s}^{(*)}$
Production at the $\Upsilon$(5S)}
\thanks{Submitted to the 32$^{\rm nd}$ International Conference on High Energy Physics,
Aug 2004, Beijing}

\author{D.~M.~Asner}
\author{S.~A.~Dytman}
\author{S.~Mehrabyan}
\author{J.~A.~Mueller}
\author{V.~Savinov}
\affiliation{University of Pittsburgh, Pittsburgh, Pennsylvania
15260}
\author{Z.~Li}
\author{A.~Lopez}
\author{H.~Mendez}
\author{J.~Ramirez}
\affiliation{University of Puerto Rico, Mayaguez, Puerto Rico 00681}
\author{G.~S.~Huang}
\author{D.~H.~Miller}
\author{V.~Pavlunin}
\author{B.~Sanghi}
\author{E.~I.~Shibata}
\author{I.~P.~J.~Shipsey}
\affiliation{Purdue University, West Lafayette, Indiana 47907}
% ...
%\author{(CLEO Collaboration)} %FOR PRD_SPECIAL_CHANGEME
\collaboration{CLEO Collaboration} %FOR PRL,CLNS
\noaffiliation

\begin{abstract}
Using data collected by the CLEO III detector at CESR, we started a
series of investigations on the $\Upsilon$(5S) resonance decay
properties. The data sample used for this analysis consists of 0.42
fb$^{-1}$ of data taken on the $\Upsilon$(5S) resonance, 6.34
fb$^{-1}$ of data collected on the $\Upsilon(4S$) and 2.32 fb$^{-1}$
of data taken in the continuum below the $\Upsilon$(4S). $B_s$
mesons are expected to decay predominantly into $D_s$ meson, while
the lighter $B$ mesons decay into $D_s$ only about 10\% of the time.
We exploit this difference to make a preliminary model dependent
estimate of the ratio of $B_s^{(*)}\overline{B_s}^{(*)}$ to the
total $b\overline{b}$ quark pair production at the $\Upsilon$(5S)
energy to be $(21\pm 3 \pm 9)$\%.
\end{abstract}
 \maketitle

\section{Introduction}
An enhancement in the total $e^+e^-$ annihilation cross-section into
hadrons was discovered at CESR long ago \cite{CLEOIImeasurement},
\cite{CUSBhyperfine} and \cite{CUSBmasses} and its mass measured as
10.865$\pm$0.008 GeV. This effect was named the $\Upsilon$(5S)
resonance. Potential models \cite{UQMwhere?}, \cite{UQMY5S} and
\cite{UQM1986} predict the different relative decay rates of the
$\Upsilon$(5S) into combinations of $B^{(*)}\overline{B}^{(*)}$ and
$B_s^{(*)} \overline{B}_s^{(*)}$ where $(*)$ indicates the possible
presence or absence of a $B^*$ meson. Some data, $\sim$ 116
pb$^{-1}$, failed to reveal if $B_s$ mesons were produced. It is
important to check the predictions of these and other models;
furthermore, $e^+e^-$ ``B factories" could exploit a possible $B_s$
yield here as they have done for $B$ mesons on the $\Upsilon$(4S).
In this paper we examine $D_s$ yields because in a simple spectator
model the $B_s$ decays into the $D_s$ nearly all the time. Since the
$B\to D_s X$ branching ratio has already been measured to be
$(10.5\pm 2.6\pm 2.5)\%$ \cite{PDG}, we expect a large difference
between the $D_s$ yields at the $\Upsilon$(5S) and the
$\Upsilon$(4S) that can lead to an estimate of the size of the
$B_s^{(*)}\overline{B_s}^{(*)}$ component at the $\Upsilon$(5S).
When we discuss the $\Upsilon$(5S) here, we mean the production of
any $B$ meson species including $B_u$, $B_d$ and $B_s$ that occurs
at an $e^+e^-$ center-of-mass energy of 10.87 GeV.

\section{The CLEO III Detector}

The CLEO III detector is well equipped to measure the momenta and
directions of charged particles, identify charged hadrons, detect
photons and measure with good precision their directions and
energies. Muons above 1.1 GeV can also be identified. The detector
is almost cylindrically symmetric with everything but the muon
detector inside a superconducting magnet coil run at a current
that produces an almost uniform 1.5 T field. The detector consists
of a four-layer double sided silicon strip detector at small
radius. It is followed by a 47-layer drift chamber that uses a gas
mixture of 60\% Helium and 40\% Propane. These two devices measure
charged track vertices and three-momenta with excellent accuracy.
The drift chamber also measures energy loss, dE/dx, that is used
to identify charged tracks below about 0.7 GeV/c. After the drift
chamber there is a Ring Imaging Cherenkov Detector (RICH)
\cite{RICH}, that separates pions from kaons from threshold up to
about 2.7 GeV/c. The RICH is surrounded by a Thallium doped CsI
crystal array consisting of about 8000 tapered crystals 30 cm long
and about 5x5 cm$^2$ at the rear.

%\section{Measurement of $D_s$ Meson Yields at the $\Upsilon$(5S) and $\Upsilon$(4S) Resonances}

\section{Data Sample and Signal Selection}

In this analysis we use 0.42 fb$^{-1}$ integrated luminosity
representing all the data taken right at the $\Upsilon$(5S) peak. We
also use 6.34 fb$^{-1}$ of integrated luminosity collected on the
$\Upsilon$(4S) and 2.32 fb$^{-1}$ of data taken in the continuum 30
MeV in center-of-mass energy below the $\Upsilon$(4S) for continuum
subtraction.

We looked for $D_s$ candidates through the reconstruction of three
charged tracks in hadronic events via the
$D_s^+\rightarrow\phi\pi^+$ decay mode. Here and elsewhere in this
paper mention of one charge implies the same consideration for the
charge-conjugate mode. Since $b$-quark events are relatively
isotropic compared to continuum background events, these latter
ones are suppressed by requiring that the Fox-Wolfram shape
parameter $R_2$ \cite{Fox} is less than $0.25$.

The tracks are required to be well measured with momenta between
$0.04$ and $3~\rm GeV$ and have at least $50\%$ of the expected
number of hits. Each track should also have distance of closest
approach to the interaction vertex in the bending plane $ \leq
0.005$ $\rm m$ and have a $z$ coordinate of the point of closest
approach in the bending ($xy$) plane less or equal to $0.05$ $\rm
m$.

We use both charged particle ionization loss in the drift chamber
(dE/dx) and RICH information to identify kaons and pions. The RICH
is used for momenta larger than 0.62 GeV. Information on the angle
of detected Cherenkov photons is translated into a Likelihood of a
given photon being due to a particular particle. Contributions
from all photons associated with a particular track are then
summed to form an overall Likelihood denoted as ${\cal L}_i$ for
each ``$i$" particle hypothesis. To differentiate between pion and
kaon candidates, we use the difference: $-2\log({\cal
L_{\pi}})+2\log({\cal L}_K$).

To utilize the dE/dx information we calculate the differences
between the expected ionization losses and the observed losses
divided by the error for the pion and kaon cases called
$\sigma_{\pi}$ and $\sigma_{K}$.

We use both the RICH and dE/dx information in the following
manner: (a) If neither the RICH nor dE/dx information is
available, then the track is accepted as both a pion and a kaon
candidate. (b) If dE/dx is available and RICH is not, then we
insist that pion candidates have
$PID_{dE}=\sigma_{\pi}^2-\sigma_{K}^2 <0$ and kaon candidates have
$PID_{dE}> 0.$ (c) If RICH information is available and dE/dx is
not available, then we require that $PID_{RICH}=-2\log({\cal
L}_{\pi})+2\log({\cal L}_K)<0$ for pions and $PID_{RICH}>0$ for
kaons. (d) If both dE/dx and RICH information are available, we
require that $(PID_{dE}+PID_{RICH}) <0$ for pions and
$(PID_{dE}+PID_{RICH})>0$ for kaons.

$D_s^+$ candidates are searched for in the
$\phi$$\pi^+$ decay mode using the $\phi\to K^+K^-$ channel. Pairs of
oppositely charged tracks were considered as candidates for the
decay products of the {$\phi$} if each track passes the previous
selection criteria (except the particle identification cut where
just one of the kaons is required to pass), and if the invariant
mass of the $K^+K^-$ system was within $\pm 10$ $\rm MeV/c^2$ of the
nominal {$\phi$} mass. A third track passing the track selection
requirements (listed above) except the particle identification cuts
%with dE/dx and RICH information consistent with the expectation
%for a pion
was combined with the $K^+K^-$ system to form a $D_s$ candidate.

To suppress combinatoric backgrounds, we take advantage of the
polarization of the $\phi$ as it is a vector particle while the
other particles are spinless. The expected distribution from real
$\phi$ decays varies as $\cos^2\theta_h$, where $\theta_h$ is the
angle between the $D_s$ and the $K^+$ momenta measured in the
$\phi$ rest frame while combinatoric backgrounds tend to be flat.
Thus, we require $|\cos\theta_h|$ to be larger than $0.3$.

\section{$D_s$ Production at the $\Upsilon$(5S) and $\Upsilon$(4S)}

\subsection{$D_s$ Mass Spectra and Yields}

\begin{figure}[htbp]
 \centerline{\epsfig{figure=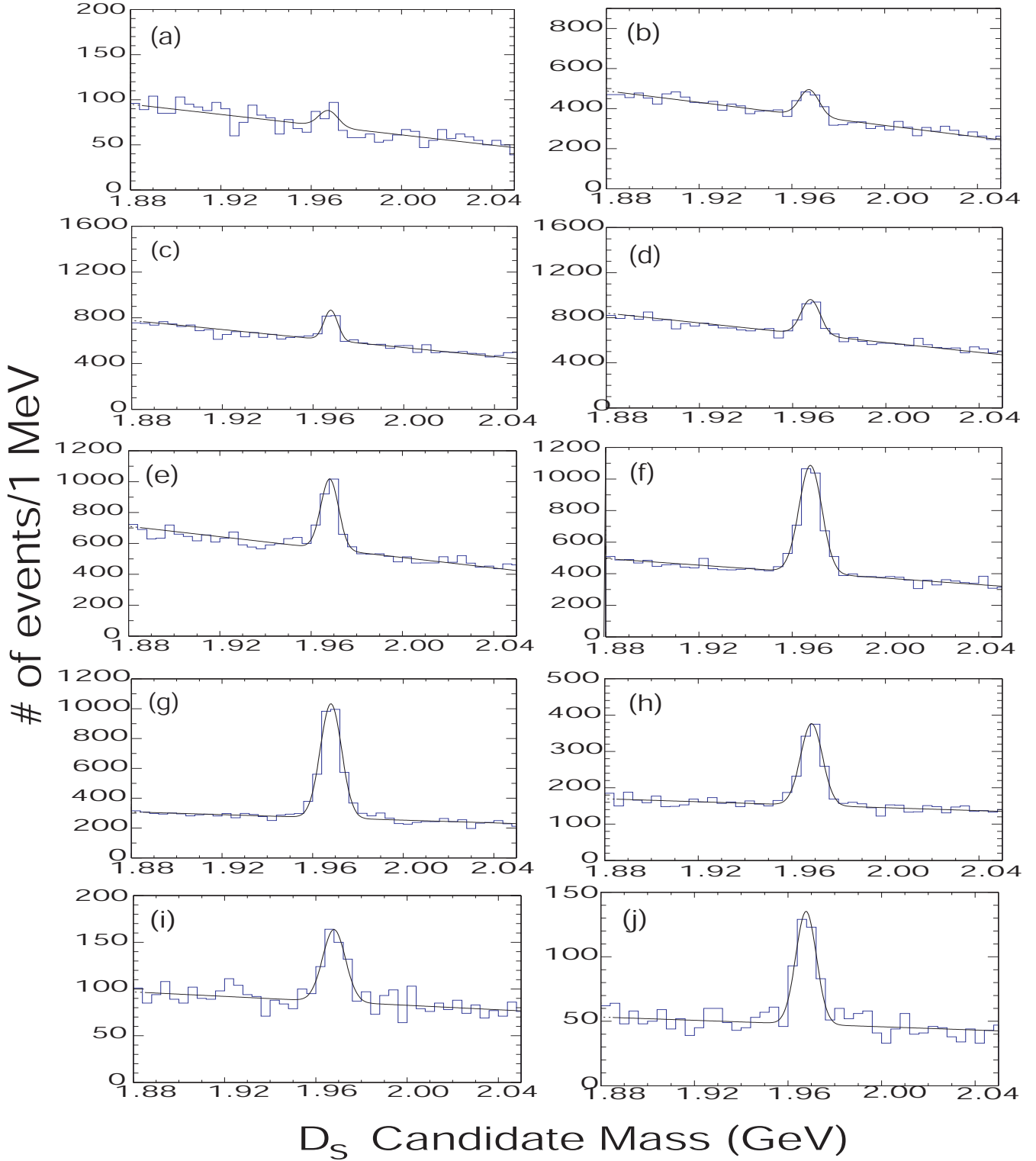,height=8in,width=7in}}
  \caption{\label{Dsmass4sonx} The $\phi\pi^+$ mass combinations, fitted to a Gaussian
  signal shape centered at the $D_s$ mass plus a polynomial
  background for $\Upsilon$(4S) on-resonance data in the $x$ intervals:
  (a) $0<x<0.05$, (b) $0.05<x<0.10$,
  (c) $0.10<x<0.15$, (d) $0.15<x<0.20$, (e) $0.20<x<0.25$, (f) $0.25<x<0.30$,
  (g) $0.30<x<0.35$, (h) $0.35<x<0.40$, (i) $0.40<x<0.45$, (j) $0.45<x<0.50$ (Preliminary).}
   \end{figure}

\begin{figure}[htbp]
 \centerline{\epsfig{figure=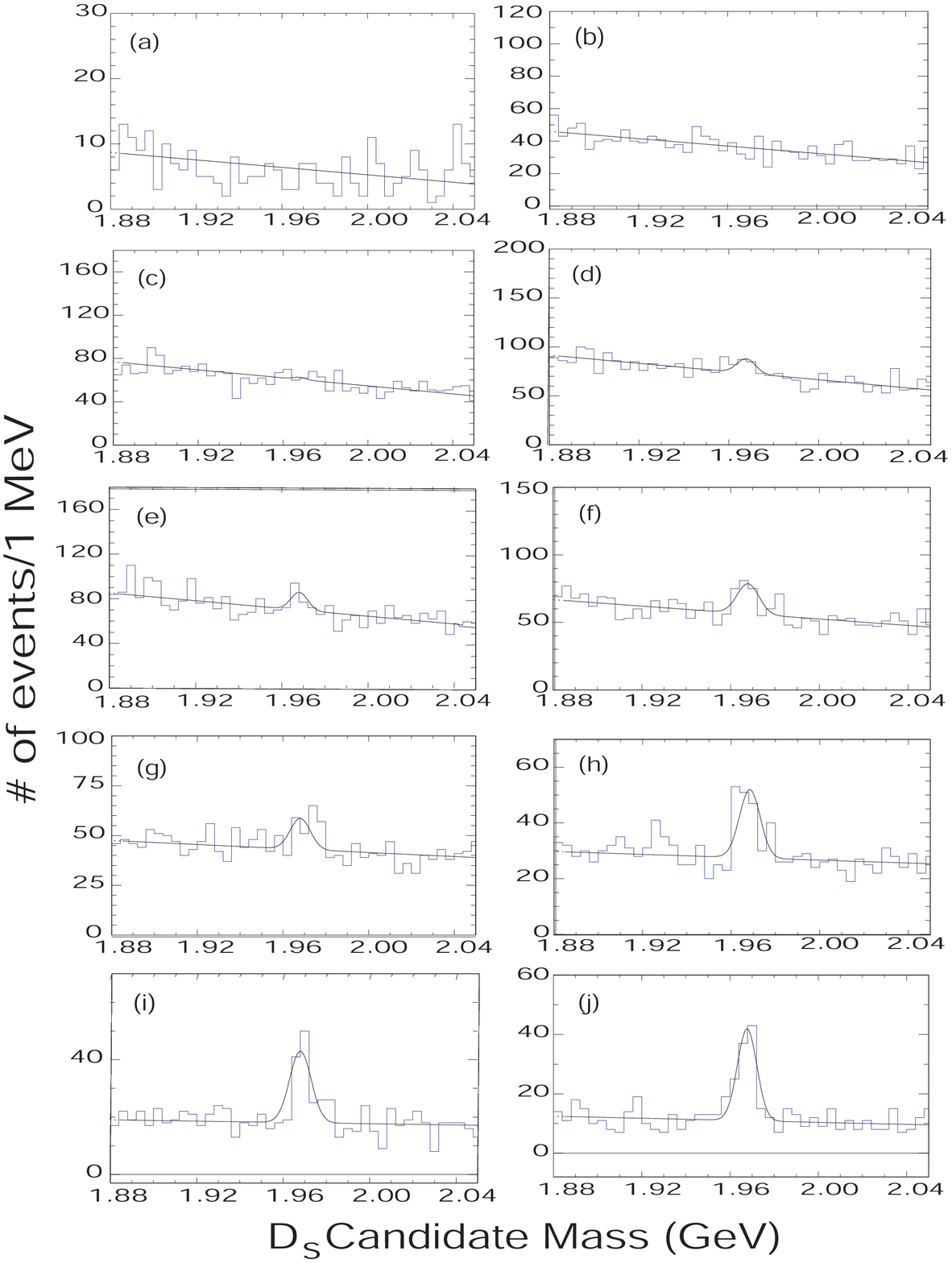,height=8in,width=7in}}
  \caption{\label{Dsmass4soffx} The $\phi\pi^+$ mass combinations, fitted to a Gaussian
  signal shape centered at the $D_s$ mass plus a polynomial
  background for the
  continuum below the $\Upsilon$(4S) data in the $x$ intervals: (a) $0<x<0.05$, (b) $0.05<x<0.10$,
  (c) $0.10<x<0.15$, (d) $0.15<x<0.20$, (e) $0.20<x<0.25$, (f) $0.25<x<0.30$,
  (g) $0.30<x<0.35$, (h) $0.35<x<0.40$, (i) $0.40<x<0.45$, (j) $0.45<x<0.50$ (Preliminary).}
   \end{figure}

\begin{figure}[htbp]
 \centerline{\epsfig{figure=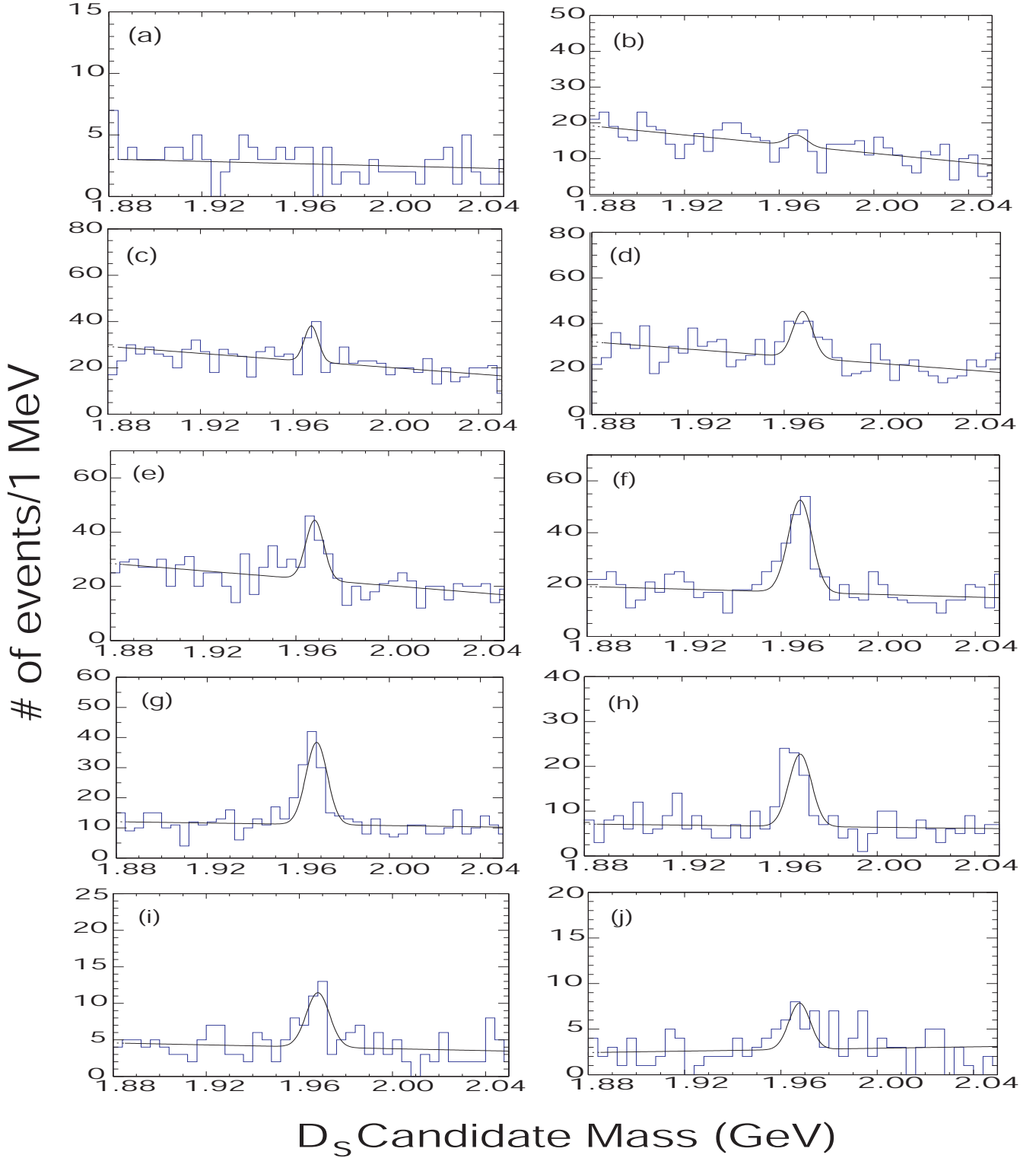,height=8in,width=7in}}
  \caption{\label{Dsmass5sx} The $\phi\pi^+$ mass combinations, fitted to a Gaussian
  signal shape centered at the $D_s$ mass plus a polynomial
  background for
  $\Upsilon$(5S) data in the $x$ intervals: (a) $0<x<0.05$, (b) $0.05<x<0.10$,
  (c) $0.10<x<0.15$, (d) $0.15<x<0.20$, (e) $0.20<x<0.25$, (f) $0.25<x<0.30$,
  (g) $0.30<x<0.35$, (h) $0.35<x<0.40$, (i) $0.40<x<0.45$, (j) $0.45<x<0.50$ (Preliminary).}
   \end{figure}

For $\phi\pi^+$ combinations satisfying the previous requirements,
we look for $D_s$ candidates having a momentum less than or equal to
half of the beam energy. Instead of momentum we choose to work with
the variable $x$ which is the $D_s$ momentum divided by the beam
energy, to remove at first order differences between continuum data
taken just below the $\Upsilon$(4S), at the $\Upsilon$(4S) and at
the $\Upsilon$(5S). Since we are interested in calculating the $D_s$
yields versus $x$, we fit the invariant mass of the $\phi\pi^{\pm}$
candidates in 10 different $x$ intervals (from 0 to 0.5) for data
taken at the $\Upsilon$(4S) peak, at the continuum below the
$\Upsilon$(4S) and at the $\Upsilon$(5S) peak as shown in
Fig.~\ref{Dsmass4sonx}, Fig.~\ref{Dsmass4soffx} and
Fig.~\ref{Dsmass5sx}, respectively.


\begin{figure}[htbp]
 \centerline{\epsfig{figure=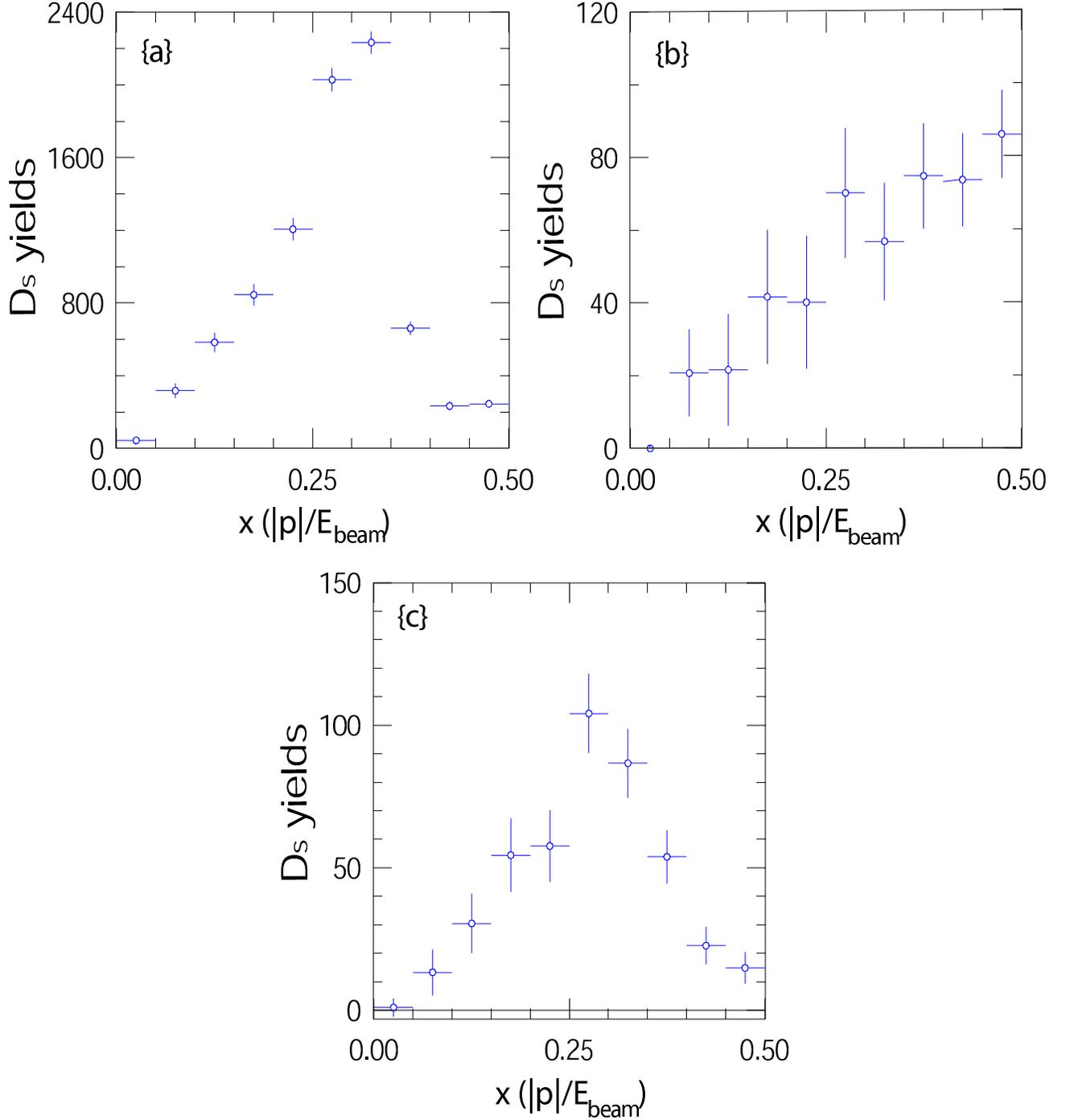,height=7.3in,width=7.3in}}
  \caption{\label{RawYields} $D_s$ yields from:
  (a) the $\Upsilon$(4S) data (b) the continuum below the $\Upsilon$(4S) data
  (c) the $\Upsilon$(5S) data (Preliminary).}
\end{figure}

Some $x$ intervals did not have enough statistics to allow floating
the fitting function parameters in each of our three data samples.
Thus we fixed the width of the Gaussian signal shapes by fitting the
large $\Upsilon$(4S) data sample in each $x$ interval allowing the
width to float. Then we fixed the widths to these values when
fitting at the other energies. The raw $D_s$ yields extracted from
the fitting are shown in Fig.~\ref{RawYields} (a), (b) and (c) and
listed in the second and third column of Table 1 and Table 2.

\subsection{Continuum Subtraction} The number of signal events is
determined by subtracting the scaled continuum data below the
$\Upsilon$(4S) from the $\Upsilon$(4S) and from the $\Upsilon$(5S)
data. We note, that the data below the $\Upsilon(4S)$ represent
four-flavor continuum events containing $u$, $d$, $s$ and $c$
quarks. To determine the scale factor, $S_i$ we used two different
methods, the first one accounts for both the ratio of luminosities
and the $s$ dependence of the continuum cross section. Here
\begin{equation}
S_i={\frac{L^{(i)}}{L_{cont}}}\cdot{ \left({\frac{E_{cont}}{E^{(i)}}}\right)^2}
\end{equation}
where $L^{(i)}$, $L_{cont}$, $E^{(i)}$ and $E_{cont}$ are the collected luminosities
and the center of mass energies at
the resonance $(i)$ and at the continuum below the $\Upsilon$(4S). We find
\begin{equation}
S_{4S}={2.713 \pm 0.001 \pm 0.029}~\label{eq:4scaleFactor11}
\end{equation}
and the scale factor between the continuum below the $\Upsilon$(4S)
and the $\Upsilon$(5S):
\begin{equation}
S_{5S}={(17.14 \pm 0.01 \pm
0.38)}\cdot{10^{-2}}~\label{eq:5scaleFactor11}
\end{equation}

The systematic errors use the absolute errors on the luminosity
determination at each energy. They are clearly conservative as part
of this error will cancel since we are concerned only with the error
in the luminosity ratio. To estimate the systematic error in an
independent manner, we do a second measurement of these scale
factors using the data and take the difference between the two
values as an estimate the systematic error. In this method we
measure the number of charged tracks between $0.5<x<0.8$ for each of
the three data sets. The ratio of tracks $\Upsilon$(4S)/Continuum
gives $S_{4S}=2.738\pm 0.008(stat)$ and for $\Upsilon$(5S)/Continuum
$S_{5S}= (16.86\pm0.04(stat))\cdot{10^{-2}}$, both with a negligible
statistical error.  These numbers differ by 1.0\% and 1.7\% from the
previous method and we take these differences as the systematic
errors. Thus we use
\begin{equation}
S_{4S}={2.713 \pm 0.001 \pm 0.026}~\label{eq:4scaleFactor1}
\end{equation}
\begin{equation}
S_{5S}={(17.14 \pm 0.01 \pm
0.28)}\cdot{10^{-2}}~\label{eq:5scaleFactor1}
\end{equation}
The four-flavor continuum subtracted $D_s$ yields at the
$\Upsilon$(4S) and $\Upsilon$(5S) are listed in the fourth column of
Tables 1 and 2, respectively.
\subsection{$D_s$ reconstruction efficiency}
\begin{figure}[htbp]
\centerline{\epsfig{figure=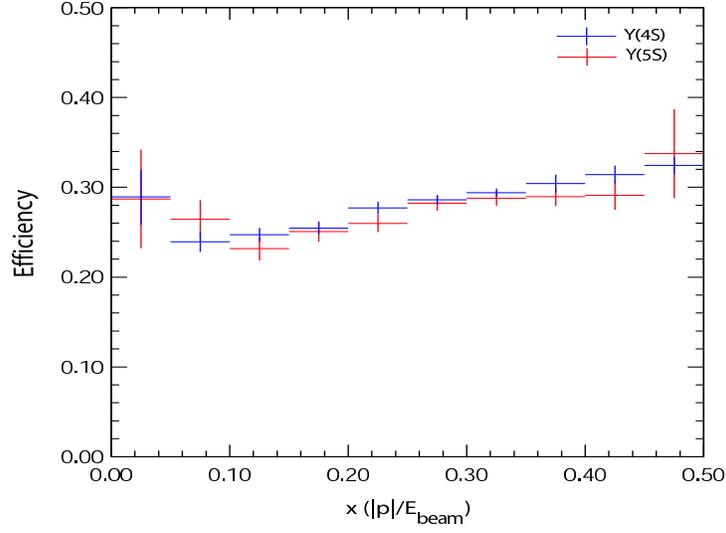,height=2.75in,width=3.75in}}
  \caption{\label{efficiency} $D_s$ reconstruction efficiency from the $\Upsilon$(4S)
  and the $\Upsilon$(5S) data(scaled).}
\end{figure}

\begin{figure}[htbp]
 \centerline{\epsfig{figure=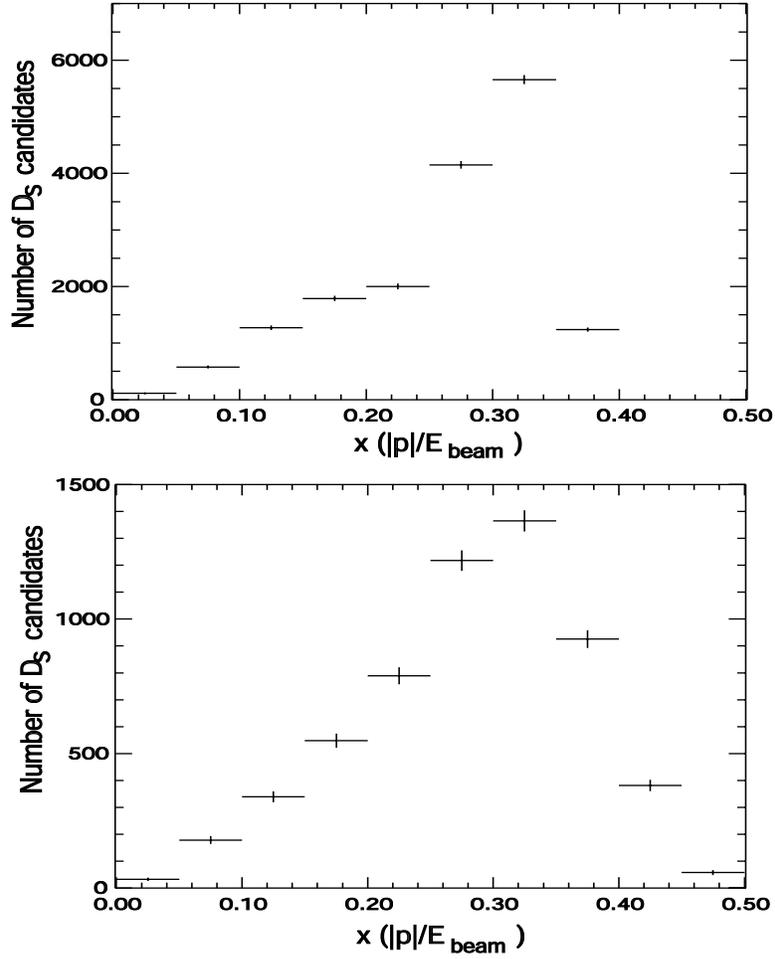,height=5in,width=4in}}
  \caption{\label{Ds_yields_mc} $D_s$ yields from:
  (a) the $\Upsilon$(4S) MC (b) the $\Upsilon$(5S) MC.}
\end{figure}

\begin{figure}[htbp]
 \centerline{\epsfig{figure=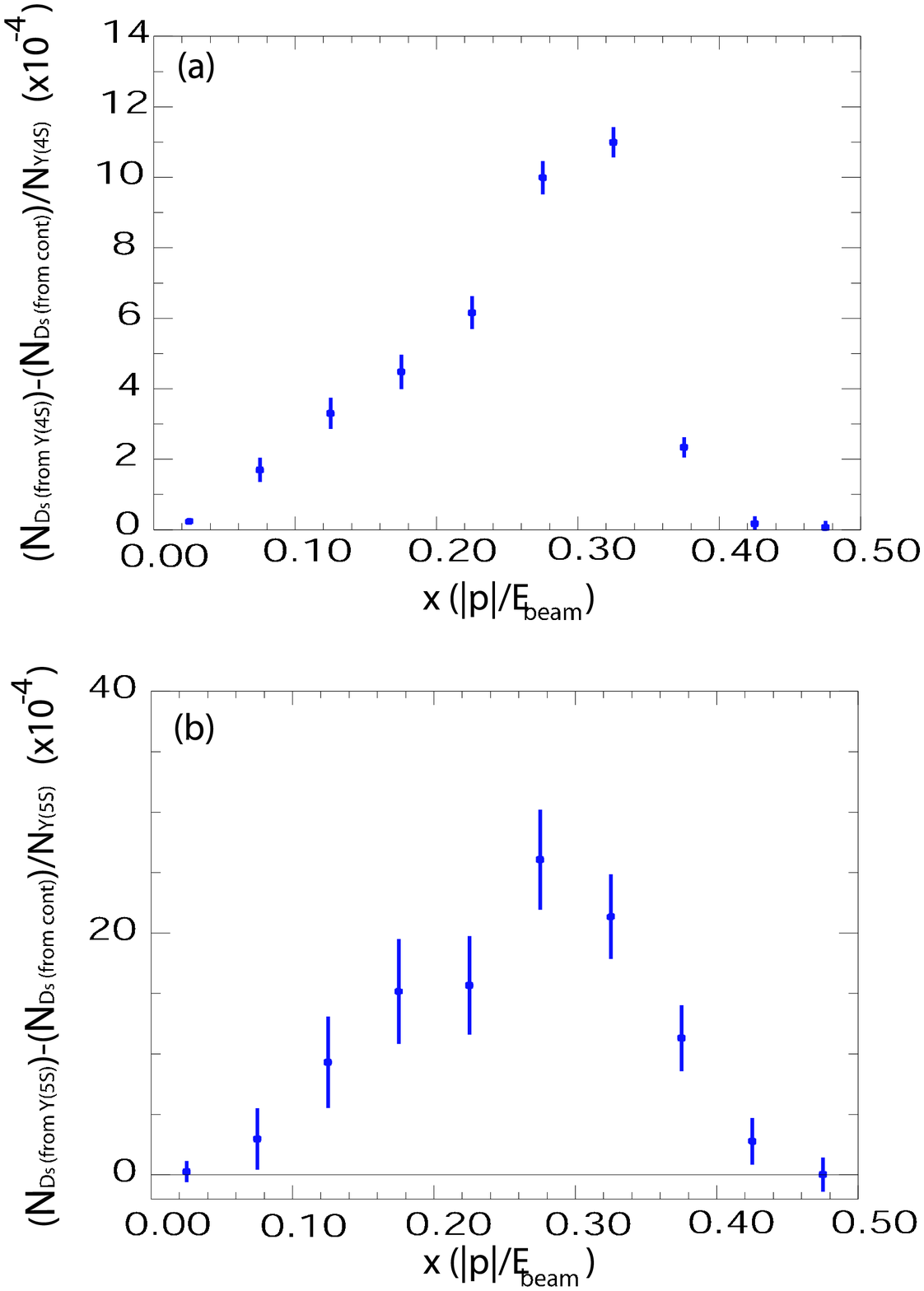,height=8in, width=5in}}
  \caption{\label{Dsyield } $D_s$ yields vs x from: (a) the $\Upsilon$(4S) decays, (b) the $\Upsilon$(5S)
  decays. Both plots are continuum subtracted, efficiency corrected, and normalized to the number of
  resonance events (Preliminary).}
\end{figure}

  In order to convert the fitted event numbers into branching ratios,
we have simulated both $\Upsilon$(4S) and $\Upsilon$(5S) $B^{(*)}$
and $B_s^{(*)}$ decays that have $D_s$ mesons in the final state.
We reconstructed the simulated $D_s$ as we did for the data and we
fitted the invariant mass to a Gaussian signal and a polynomial
background shape. The $x$ dependent $D_s$ detection efficiencies
at the $\Upsilon$(4S) and $\Upsilon$(5S) are shown in
Fig.~\ref{efficiency}. The efficiencies of two datasets are
consistent with each other and since there is no reason to believe
that they differ, in what follows we assume they are equal.
The $x$ dependent $D_s$ yields extracted from the fit of the
$\Upsilon$(4S) and $\Upsilon$(5S) simulated events are shown in
Fig.~\ref{Ds_yields_mc} (a) and (b).
We show in Fig.~\ref{Dsyield }(a) and (b) the $x$ distribution of
the inclusive $D_s$ yields from the $\Upsilon$(4S) and the
$\Upsilon$(5S) decays respectively, continuum subtracted, efficiency
corrected, and normalized to the number of $\Upsilon$(4S) and
$\Upsilon$(5S) resonance events.

\subsection{$D_s$ production rates at the $\Upsilon$(4S) and the
$\Upsilon$(5S)}

The total number of hadronic events above four-flavor continuum at
the energies corresponding to the $\Upsilon$(4S) and $\Upsilon$(5S)
energies are

\begin{equation}
N^{Res}_{\Upsilon(4S)}=N^{on}_{\Upsilon(4S)}-S_{4S}*N^{off}_{\Upsilon(4S)}=6,420,910\pm
5,738\pm 240,542~\label{eq:RSol1}
\end{equation}
\begin{equation}
N^{Res}_{\Upsilon(5S)}=N^{on}_{\Upsilon(5S)}-S_{5S}*N^{off}_{\Upsilon(5S)}=131,396\pm
810 \pm 26,546~\label{eq:RSol2}
\end{equation}
Using these numbers together with $N^{i}_{\Upsilon(4S)}$
($N^{i}_{\Upsilon(5S)}$) and $\epsilon^{i}$ which are the $D_s$
yield and $D_s$ reconstruction efficiency in the $i$-th $x$
interval, we measure the partial $\Upsilon$(4S) and $\Upsilon$(5S)
to $D_s X$ branching ratios in the ith $x$ interval as follows

\begin{eqnarray}
{\cal{B}}^{i}(\Upsilon(4S)\to
D_sX)&=&{\frac{1}{N^{Res}_{\Upsilon(4S)}*{{\cal{B}}(D_{s}\to \phi
\pi)}*{{\cal{B}}(\phi\to K^+K^-)}}}~*\left({\frac{N^{i}_{\Upsilon(4S)}}
{\epsilon^{i}}}\right)~\label{eq:RSol4}\\
{\cal{B}}^{i}(\Upsilon(5S)\to
D_sX)&=&{\frac{1}{N^{Res}_{\Upsilon(5S)}*{{\cal{B}}(D_{s}\to \phi
\pi)}*{{\cal{B}}(\phi\to K^+K^-)}}}~*\left({\frac{N^{i}_{\Upsilon(5S)}}
{\epsilon^{i}}}\right)~~.\label{eq:RSol5}
\end{eqnarray}

The results are listed in Tables 2 and 3 respectively.

\begin{table}[htb]
\begin{center}
\begin{tabular}{cccccc}
$x^{i}$($\frac{|p|}{Ebeam}$)~ &~ ON $\Upsilon$(4S) & Continuum &
$N^{i}_{\Upsilon(4S)}$  & $\epsilon^{i}(\%)$ & $B^{i}$(\%)\\
 \hline
0.00-0.05 & $44.4\pm15.7$   &~ $0.0\pm0.0$   &~  $44\pm16 $   &~  $28.9 $  &   $0.1\pm0.1 $\\
0.05-0.10 & $317.6\pm39.6$  &~ $20.7\pm12.0$ &~  $261\pm51 $  &~  $23.9 $  &   $1.0\pm0.2 $\\
0.10-0.15 & $583.6\pm53.9$  &~ $21.6\pm15.3$ &~  $525\pm68 $  &~  $24.7 $  &   $1.9\pm0.5 $\\
0.15-0.20 & $845.5\pm59.0$  &~ $41.7\pm18.5$ &~  $732\pm77 $  &~  $25.4 $  &   $2.5\pm0.6 $\\
0.20-0.25 & $1206.4\pm60.6$ &~ $40.2\pm18.3$ &~  $1097\pm78 $ &~  $27.7 $  &   $3.5\pm0.9 $\\
0.25-0.30 & $2028.6\pm63.8$ &~ $70.3\pm18.0$ &~  $1838\pm80 $ &~  $28.6 $  &   $5.6\pm1.4 $\\
0.30-0.35 & $2233.7\pm60.7$ &~ $57.0\pm16.2$ &~  $2079\pm75 $ &~  $29.4 $  &   $6.2\pm1.6 $\\
0.35-0.40 & $660.8\pm37.9$  &~ $75.0\pm14.5$ &~  $457\pm55 $  &~  $30.4 $  &   $1.3\pm0.3 $\\
0.40-0.45 & $233.5\pm25.9$  &~ $73.4\pm12.9$ &~  $34\pm43 $   &~  $31.4 $  &   $0.1\pm0.1 $\\
0.45-0.50 & $245.8\pm22.2$  &~ $86.0\pm12.1$ &~  $13\pm40 $   &~  $32.4 $  &   $0.0\pm0.0 $\\
\hline\hline
\end{tabular}
\end{center}
\caption{$D_s$ yields from the $\Upsilon$(4S) data, continuum below
the $\Upsilon$(4S) and the $\Upsilon$(4S) continuum subtracted data
($N^{i}_{\Upsilon(4S)}$). Also listed are the $D_s$ reconstruction
efficiencies ($\epsilon^{i}$), and the partial $\Upsilon(4S)\to D_s
X$ branching fractions vs $x$ (Preliminary).} \label{tab:Drecon}
\end{table}

\begin{table}[htb]
\begin{center}
\begin{tabular}{cccccc}
   $x^{i}$($\frac{|p|}{Ebeam}$) & ON $\Upsilon$(5S) &
Continuum & $N^{i}_{\Upsilon(5S)}$  &  $\epsilon^{i}(\%)$  &  $B^{i}$(\%)  \\
    \hline
0.00-0.05 & $1.0\pm3.2$    &~ $0.0\pm0.0$    &~ $1\pm3$       &~ $28.9$  &  $0.1\pm0.1 $\\
0.05-0.10 & $13.3\pm8.1$   &~ $20.7\pm12.0$  &~ $9.7\pm8.3$   &~ $23.9$  & $1.8\pm1.6 $\\
0.10-0.15 & $30.4\pm10.4$  &~ $21.6\pm15.3$  &~ $26.7\pm10.7$ &~ $24.7$  & $4.7\pm2.2 $\\
0.15-0.20 & $54.4\pm13.0$  &~ $41.7\pm18.5$  &~ $47.2\pm13.3$ &~ $25.4$  & $8.0\pm3.0 $\\
0.20-0.25 & $57.6\pm12.7$  &~ $40.2\pm18.3$  &~ $50.7\pm13.0$ &~ $27.7$  & $7.9\pm2.8 $\\
0.25-0.30 & $104.1\pm14.0$ &~ $70.3\pm18.0$  &~ $92.0\pm14.3$ &~ $28.6$  & $13.9\pm4.1 $\\
0.30-0.35 & $86.7\pm12.1$  &~ $57.0\pm16.2$  &~ $76.9\pm12.4$ &~ $29.4$  & $11.3\pm3.4 $\\
0.35-0.40 & $53.9\pm9.4$   &~ $75.0\pm14.5$  &~ $41.0\pm9.7$  &~ $30.4$  & $5.8\pm2.0 $\\
0.40-0.45 & $22.6\pm6.7$   &~ $73.4\pm12.9$  &~ $10.1\pm7.0$  &~ $31.4$  & $1.4\pm1.0 $\\
0.45-0.50 & $14.9\pm5.6$   &~ $86.0\pm12.1$  &~ $0.1\pm6.0$   &~ $32.4$  & $0.0\pm0.8 $\\
\hline\hline
\end{tabular}
\end{center}
\caption{$D_s$ yields from the $\Upsilon$(5S) data, continuum below
the $\Upsilon$(4S) and $\Upsilon$(5S) continuum subtracted
data($N^{i}_{\Upsilon(5S)}$). Also listed are the $D_s$
reconstruction efficiencies($\epsilon^{i}$), and the partial
$\Upsilon(5S) \to D_s X$ branching fractions vs $x$ (Preliminary).}
\label{tab:Drecon}
\end{table}

To compute the total production rate, we sum the partial production
rates to obtain
\begin{eqnarray}
{\cal{B}}(\Upsilon(4S)\to
D_sX)&=&{\frac{1}{N^{Res}_{\Upsilon(4S)}*{{\cal{B}}(D_{s}\to \phi
\pi)}*{{\cal{B}}(\phi\to K^+K^-)}}}\sum_{i}\left({\frac{N^{i}_{\Upsilon(4S)}}
{\epsilon^{i}}}\right)~\label{eq:RSol4}\\
{\cal{B}}(B\to D_sX)&=&{\frac{1}{2}}*{{\cal{B}}(\Upsilon(4S)\to D_sX)}~\label{eq:RSol6}\\
{\cal{B}}(\Upsilon(5S)\to
D_sX)&=&{\frac{1}{N^{Res}_{\Upsilon(5S)}*{{\cal{B}}(D_{s}\to \phi
\pi)}*{{\cal{B}}(\phi\to K^+K^-)}}}\sum_{i}\left({\frac{N^{i}_{\Upsilon(5S)}}
{\epsilon^{i}}}\right)~~.\label{eq:RSol5}
\end{eqnarray}

Therefore, we measure the following preliminary quantities. The
product of the $D_s$ production rate at the $\Upsilon$(4S) and the
${\cal{B}}(D_s\to\phi\pi)$ is
\begin{equation}
{\cal{B}}(\Upsilon(4S)\to D_sX)\cdot{\cal{B}}(D_s\to\phi\pi)=
{\frac{\sum_{i}\left({\frac{N^{i}_{\Upsilon(4S)}}
{\epsilon^{i}}}\right)}{N^{Res}_{\Upsilon(4S)}*{{\cal{B}}(\phi\to
KK)}}}=(8.0\pm 0.3 \pm 0.4)\cdot{10^{-3}}~\label{eq:rate4}
\end{equation}

and the product of the $D_s$ production rate at the $\Upsilon$(5S)
and the ${\cal{B}}(D_s\to\phi\pi)$ is
\begin{equation}
{\cal{B}}(\Upsilon(5S)\to D_sX)\cdot{\cal{B}}(D_s\to\phi\pi)=
{\frac{\sum_{i}\left({\frac{N^{i}_{\Upsilon(5S)}}
{\epsilon^{i}}}\right)}{N^{Res}_{\Upsilon(5S)}*{{\cal{B}}(\phi\to
KK)}}}=(20\pm 2 \pm 4)\cdot{10^{-3}}~~,\label{eq:rate5}
\end{equation}
thus demonstrating a much larger production of $D_s$ at the $\Upsilon$(5S)
energy than at the $\Upsilon$(4S).

Using ${\cal{B}}(D_s\to\phi\pi^+)= ( 3.6 \pm 0.9 )\%  $  \cite{PDG},
we find
\begin{equation}
{\cal{B}}(\Upsilon(4S)\to D_sX)=(22.3\pm 0.7\pm
5.7)\%~~\label{eq:4stoDsMeasured}
\end{equation}

hence:
\begin{equation}
{\cal{B}}(B\to D_sX)=(11.1\pm0.4\pm2.9)\%~~,\label{eq:BtoDsMeasured}
\end{equation}
which is in a good agreement with the PDG \cite{PDG} value of
\begin{equation}
{\cal{B}}(B\to D_sX)=(10.5\pm 2.6 \pm 2.5)\%~~.\label{eq:BtoDsPDG}
\end{equation}
In addition, we find
\begin{equation}
{\cal{B}}(\Upsilon(5S)\to D_sX)=(55.0\pm 5.2 \pm
17.8)\%~.~\label{eq:5stoDsMeasured}
\end{equation}

\section{$B_s$ production at the $\Upsilon$(5S)}

\begin{figure}[htbp]
 \centerline{\epsfig{figure=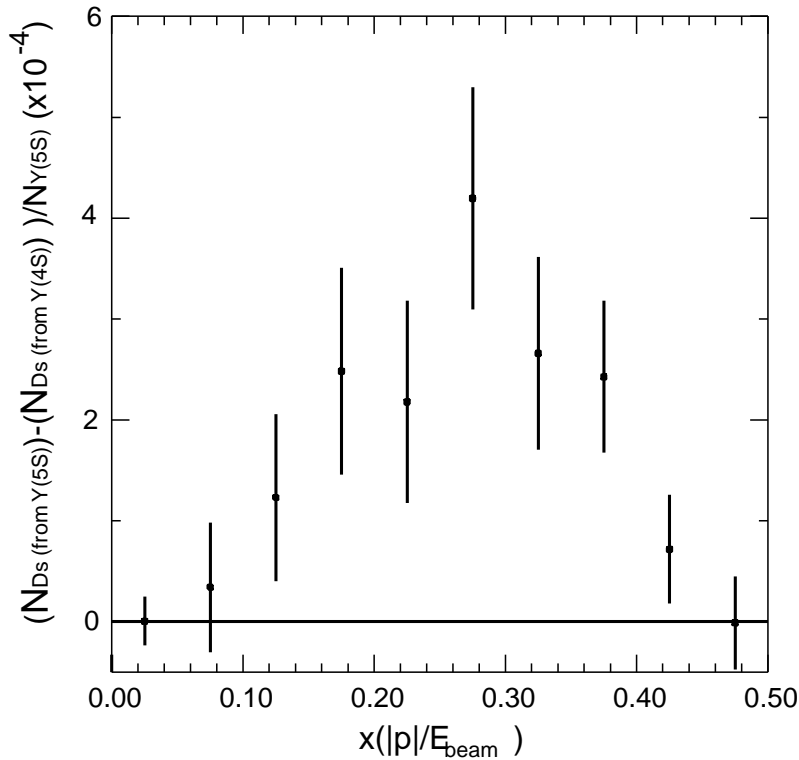,height=4in}}
  \caption{\label{enhancement} The enhancement of $D_s$ yields at the
$\Upsilon$(5S) vs $x$ (no efficiency correction, Preliminary).}
\end{figure}

We want to see if $D_s$ production at the $\Upsilon$(5S) is in
excess of what is expected from $B\overline{B}$ alone. In
Fig.~\ref{enhancement} we show the $D_s$ $x$ spectrum from the
$\Upsilon$(5S) with the $\Upsilon$(4S) spectrum subtracted. The data
were normalized such that the number of continuum subtracted
resonance decay events were the same in both cases. The spectrum
shows a significant excess of $D_s$ at the $\Upsilon$(5S), which is
a significant evidence for $B_s$ production at the $\Upsilon$(5S).

From these results, we can estimate the size of
$B_s^{(*)}\overline{B}_s^{(*)}$ component at the $\Upsilon$(5S) in a
model dependent manner.
\begin{figure}[htbp]
 \centerline{\epsfig{figure=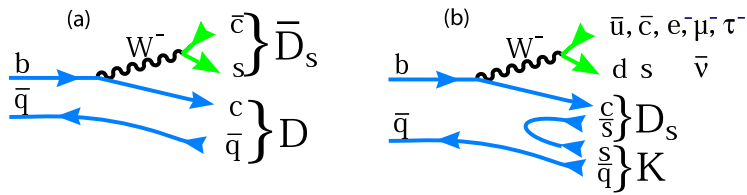,height=1in}}
  \caption{\label{BtoDs} Dominant decay diagrams for a B meson into $D_s$ mesons
  ($q$ can be here either a $u$ or a $d$ quark).}
\end{figure}
We know that the preliminary $B\to D_sX$ branching ratio we measure
\begin{equation}
{\cal{B}}(B\to D_sX)=(11.1\pm 0.4\pm 2.9)\%~\label{eq:BtoDsMeasured}
\end{equation}
comes either from the the $W^-\to \overline {c}s$ or from the $b\to
c$ piece if it manages to create an $s\overline{s}$ pair through
fragmentation as shown in Fig.~\ref{BtoDs}(a) and
Fig.~\ref{BtoDs}(b) respectively.

\begin{figure}[htbp]
 \centerline{\epsfig{figure=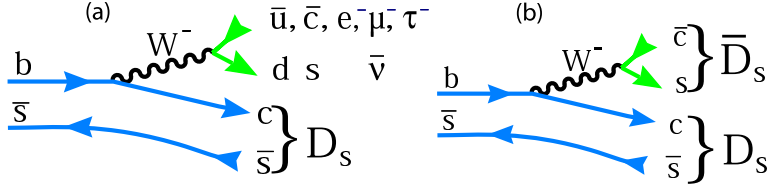,height=1in}}
  \caption{\label{BstoDs} Dominant decay diagrams for a $B_s$ meson into $D_s$ mesons.}
\end{figure}

Similarly, the production of $D_s$ mesons from $B_s$ decay arises
from two dominant processes as well. Fig.~\ref{BstoDs} shows a
large, possible greater than $100\%$ $D_s$ rate; here the primary
$b\to c$ transition has the charm quark pairing with the spectator
anti-strange quark. $D_s$ can also be produced from the upper
vertex in Fig.~\ref{BstoDs}(a) when the $W^-\to \overline {c}s$
and these two quarks form a color singlet pair. The chances of
this occurring should be similar to the chance of getting an
upper-vertex $D_s$ in $B$ decay (Fig.\ref{BtoDs}), that is, a
$D_s$ along with a $D$. We can use data to give us a guide to
these processes. The $B\to D D_s$ modes have branching fractions
that sum to about $5\%$. The observations of such $B$ decays
together with the consideration of some additional decays due to
both measured and unmeasured $B\to D^{**} D_s$ and $B\to D
D_{sJ}^{(*)}$ decays, lead to an estimate of an extra ($7\pm3)\%$
of $D_s$ mesons in $B_s$ decays into $D_s^{+(*)}D_s^{-(*)}$ final
state. However, It is possible that some $D_s$ are lost from these
processes since any $c\overline{s}$ pairs in Fig.\ref{BstoDs}(a)
and Fig.\ref{BstoDs}(b) could fragment into a kaon with a $D$
particle instead of a $D_s$ by producing a $u\overline{u}$ or
$d\overline{d}$. We don't actually know the size of the
fragmentation though it's clear that producing a light
quark-antiquark pair ($d\overline{d}$ or $u\overline{u}$) is
easier than $s\overline{s}$. Therefore, reducing the yield from
the $b\to c$ transition due to fragmentations is estimated to be a
$(-15\pm10)\%$ effect. Thus we get a model dependent estimate of
$(100+7-15)\%=92\%$,
%[I would have said $(100+7)(1-.15)=91$

\begin{equation}
{\cal{B}}(B_s\to D_sX)=(92\pm11)\%~\label{eq:BstoDsModel}
\end{equation}

We can estimate now the fraction of the $\Upsilon$(5S) that decays
into $B_s^{(*)}\overline{B}_s^{(*)}$, which we denote as $f_s$.
The $D_s$ yields at the $\Upsilon(5S)$ comes from two sources, $B$
and $B_s$ mesons. The equation linking them is

\begin{eqnarray}
{\cal{B}}(\Upsilon(5S)\to D_s X) {\cal{B}}(D_s\to\phi\pi^+)/2=
f_s\cdot {\cal{B}}(B_s\to D_sX){\cal{B}}
(D_s\to\phi\pi^+)\nonumber\\+\frac{(1-f_s)}{2}
\cdot{{\cal{B}}(\Upsilon(4S)\to D_s X)} {\cal{B}}(D_s\to\phi\pi^+)~~
\end{eqnarray}

Where the product branching fractions ${\cal{B}}(\Upsilon(5S)\to D_s
X).{\cal{B}}(D_s\to\phi\pi^+)$ and ${\cal{B}}(\Upsilon(4S)\to D_s
X).{\cal{B}}(D_s\to\phi\pi^+)$ are given by equations \ref{eq:rate5}
and \ref{eq:rate4} respectively. Therefore, we obtain the
preliminary estimate of the $B_s^{(*)}\overline{B_s}^{(*)}$ ratio to
the total $b\overline{b}$ quark pair production at the
$\Upsilon$(5S) energy
\begin{eqnarray}
f_s&=&{\cal{B}}(\Upsilon(5S)\to
B_s^{(*)}\overline{B}_s^{(*)})=(21\pm 3 \pm 9)
\%~\label{eq:5stoBsEstimated}
\end{eqnarray}

 Our result agrees with the theoretical expectations \cite{UQMwhere?} and
\cite{UQMY5S}.

\section{Sources and Estimation of Systematic Errors}
The systematic errors in this analysis
have a large component due to the $25\%$ error on the absolute $D_s\to\phi\pi$
branching fraction.

There is also a  component from the $D_s$ detection efficiency of
4.1\%, which includes a $2\%$ error on the tracking efficiency and a
$2\%$ error on the particle identification, both per track. We also
have $5\%$ error on the yields due to the fitting method.

The $1\%$ relative
error on $S_{4S}$ and $1.7\%$ on $S_{5S}$ scale factors also contribute
large components to the error on the number of hadronic events above continuum
at the $\Upsilon$(4S) and $\Upsilon$(5S) energies. Work will continue to
improve the errors.

The total systematic error is obtained by summing all entries in
quadrature.

\section{Conclusions}
We present here the first preliminary evidence of a substantial
production of $B_s$ mesons at the $\Upsilon$(5S) resonance. Using
a model dependent estimate of ${\cal{B}}(B_s\to D_s X)$, we
determine the $\Upsilon(5S) \to B_s^{(*)}\overline{B}_s^{(*)}$
branching fraction.

There have been several published phenomenological predictions of
the different relative decay rates of the $\Upsilon$(5S) into
combinations of $B^{(*)}\overline{B}^{(*)}$ and $B_s^{(*)}
\overline{B}_s^{(*)}$. The hadronic cross section in the Upsilon
region is fairly well described by the Unitarized quark model
(UQM) \cite{UQMwhere?} and \cite{UQMY5S} which is a coupled
channel model. In the Upsilon region, the model predicts the
$b$-quark state meson cross section to be dominated by
$B^{(*)}\overline{B}^{(*)}$ and $B_s^{(*)} \overline{B}_s^{(*)}$
production, with the latter one accounting for about one third of
the total $\Upsilon$(5S) cross section. However other models exist
that predict a smaller amount of $B_s$ at the $\Upsilon$(5S).

Using 131,396 $\Upsilon$(5S) decays and 6.42 million
$\Upsilon$(4S) decays collected by the CLEO III detector, we have
started a series of investigations to open the mysteries of the
$\Upsilon$(5S) properties by performing the inclusive $D_s$ meson
study, from which we present the preliminary measurements:

\begin{itemize}
\item We presented the $D_s$ meson $x$ dependent mass and yield
spectra at the continuum below the $\Upsilon$(4S) resonance, at the
$\Upsilon$(4S) peak and at the $\Upsilon$(5S) peak.
\end{itemize}

\begin{itemize}
\item From the $\Upsilon$(4S) data, we measured the production rate of $D_s$
mesons at the $\Upsilon$(4S) to be:
\begin{equation}
{\cal{B}}(\Upsilon(4S)\to D_sX)\cdot{\cal{B}}(D_s\to\phi\pi)=(8.0\pm
0.3 \pm 0.4)\times{10^{-3}}~
\end{equation}
and using ${\cal{B}}(D_s\to\phi\pi^+)= ( 3.6 \pm 0.9 )\%  $
\cite{PDG} we find
\begin{equation}
{\cal{B}}(\Upsilon(4S)\to D_sX)=(22.3\pm 0.7 \pm 5.7)\%~
\end{equation}
and hence the Branching ratio:
\begin{equation}
{\cal{B}}(B\to D_sX)=(11.1\pm 0.4\pm 2.9)\%~ \end{equation}
\end{itemize}

\begin{itemize}
\item From the $\Upsilon$(5S) data, we measured the production rate of $D_s$
mesons at the $\Upsilon$(5S), which has never been measured before to be:
\begin{equation}
{\cal{B}}(\Upsilon(5S)\to
D_sX)\cdot{\cal{B}}(D_s\to\phi\pi)=(2.0\pm 0.2 \pm
0.4)\times{10^{-2}}~
\end{equation}
Using again ${\cal{B}}(D_s\to\phi\pi^+)= ( 3.6 \pm 0.9 )\%  $
\cite{PDG}, we find
\begin{equation}
{\cal{B}}(\Upsilon(5S)\to D_sX)=(55.0\pm 5.2 \pm 17.8)\%~
\end{equation}
\end{itemize}

\begin{itemize}
\item We compared the $D_s$ production rates at the $\Upsilon$(5S)
with the $\Upsilon$(4S) and we found
\begin{equation}
\frac{{\cal{B}}(\Upsilon(5S)\to D_s X)}{{\cal{B}}(\Upsilon(4S)\to
D_s X)}=2.5\pm 0.3\pm 0.6~\label{eq:RSol9}
\end{equation}

Using ${\cal{B}}(B_s \to D_s X)=( 92\pm 11)\%$, we demonstrate a
substantial, model dependent estimate of ratio of
$B_s^{(*)}\overline{B_s}^{(*)}$ to the total $b\overline{b}$ quark
pair production at the $\Upsilon$(5S) energy

\begin{equation}
{\cal{B}}(\Upsilon(5S)\to B_s^{(*)}\overline{B}_s^{(*)})=(21\pm 3
\pm 9 )\%~
\end{equation}

\end{itemize}

Our result, at the current level of precision, is consistent with
the previously published phenomenological prediction cited above.

\end{document}